\documentclass[fleqn, twoside]{article}
\usepackage{graphics}
\begin{document}
\title{Gravitational fields of rotating disks and black holes}
\author{R.\ Meinel\\ University of Jena, 
Institute of Theoretical Physics,\\  
Max-Wien-Platz  1, 07743 Jena, Germany} 
\date{meinel@tpi.uni-jena.de}
\maketitle
\begin{abstract}
The two known exact solutions of Einstein's field equations describing
rotating objects of physical significance -- a black hole 
and a rigidly rotating
disk of dust  -- 
are discussed using a single mathematical framework related to
Jacobi's inversion problem. Both solutions can be represented in such a form
that they differ in the choice of a complex parameter and a real solution of
the axisymmetric Laplace equation only. 

A recently found family of solutions describing 
{\it differentially}
rotating disks of dust fits into the
same scheme.
\end{abstract}
\section{Introduction}
\label{intro}
Infinitesimally thin disks and black holes can be treated by 
means of the vacuum
Einstein equations. In both cases boundary value problems are to be solved, 
cf.~the contribution by G.~Neugebauer \cite{neu1}. 

In the stationary and axially 
symmetric case the vacuum Einstein equations are equivalent to the Ernst
equation
\begin{equation}
(\Re f)\triangle f = (\nabla f)^2
\label{ernst}
\end{equation}
with $\triangle = \frac{\partial^2}{\partial\varrho^2}+\frac{1}{\varrho}
\frac{\partial}{\partial\varrho}+\frac{\partial^2}{\zeta^2}$ and $\nabla =
(\frac{\partial}{\partial\varrho}, \frac{\partial}{\partial\zeta})$, where
$\varrho$ and $\zeta$ are cylindrical (Weyl-)
coordinates. (The $\zeta$-axis represents
the axis of symmetry.) The full
metric can be calculated from the complex Ernst potential $f(\varrho,\zeta)$.

By means of soliton-theoretical techniques it was possible to 
solve the problem
of a rigidly rotating disk of dust in terms of ultraelliptic functions 
\cite{nm95, nkm}. The mathematical structure of this solution allowed a 
generalization
to a class of solutions related to Jacobi's inversion problem in the general  
(hyperelliptic) case \cite{mn96}.
These solutions turned out to be closely related to finite-gap solutions
of the Ernst equation \cite{kor89, kor93, kor97, mn97}. 

In this paper I will
discuss a subclass of solutions related to the ultraelliptic case of
Jacobi's inversion problem. They contain the Kerr solution  
describing a rotating black hole, the above mentioned solution found by 
Neugebauer and
Meinel \cite{nm95} describing
a rigidly rotating disk of dust, and a three-parameter
family of solutions recently found by Ansorg and Meinel \cite{ans} 
describing differentially rotating disks of dust. In this formulation, the
solutions differ in the choice  of a complex parameter and a real solution of
the axisymmetric Laplace equation only. This will provide further insight 
into a 
certain parameter limit (``ultrarelativistic limit'') where the disk solutions
coincide with the extreme Kerr solution.
\section{Solutions of the Einstein equations 
related to Jacobi's inversion problem}
\begin{figure}
\scalebox{0.7}{\includegraphics{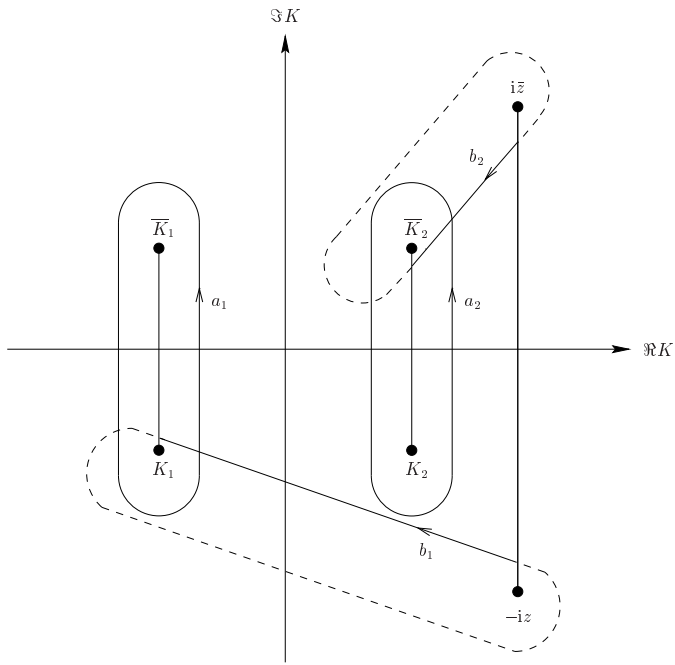}}
\caption{The $a$- and $b$-periods. Dotted lines are on the lower sheet
of the Riemann surface.}
\label{fig1}
\end{figure}
In \cite{mn96} it has been shown that 
\begin{equation}
f=\exp\left(\,\,\int\limits_{K_1}^{K_a}\frac{K^2dK}{Z} + 
\int\limits_{K_2}^{K_b}\frac{K^2dK}{Z} -v_2\right)
\label{f}
\end{equation}
with
\begin{equation}
Z=\sqrt{(K-K_1)(K-\overline{K}_1)(K-K_2)(K-\overline{K}_2)(K+{\rm i}z)
(K-{\rm i}\bar{z})},
\label{Z}
\end{equation}
a bar denoting complex conjugation, $K_1$ and $K_2$ being 
arbitrary complex parameters, and
\begin{equation}
z=\varrho+{\rm i}\zeta,
\end{equation}
represents a solution of the Ernst equation (\ref{ernst}) if 
$K_a$ and $K_b$ (and the integration paths)
are determined from Jacobi's inversion problem 
\begin{equation}
\int\limits_{K_1}^{K_a}\frac{dK}{Z} + \int\limits_{K_2}^{K_b}\frac{dK}{Z} =
v_0, \quad \int\limits_{K_1}^{K_a}\frac{KdK}{Z} + 
\int\limits_{K_2}^{K_b}\frac{KdK}{Z} = v_1,
\label{j}
\end{equation}
where $v_0$ is an arbitrary real solution of the (axisymmetric) Laplace
equation $\triangle v_0 = 0$ and the real functions 
$v_1$ and $v_2$ satisfy the differential 
relations
\begin{equation}
{\rm i}v_{j,z}=\frac{1}{2}v_{j-1}+z v_{j-1,z}, \quad j=1,2.
\label{rec}
\end{equation}
(As a consequence, $v_1$ and $v_2$ are solutions of the Laplace equation as 
well.)
The ultraelliptic functions $K_a(v_0,v_1)$ and $K_b(v_0,v_1)$ have four 
independent periods corresponding to the closed integrals in the two-sheeted
Riemann surface related to (\ref{Z}) as indicated in  Fig.\ \ref{fig1}. 
They are
called $a$- and $b$-periods according to the integration contours $a_1$, 
$a_2$, $b_1$, and $b_2$.

Sometimes the following reformulation of Eqs.\ (\ref{f}), (\ref{j}) proves to 
be useful: 
\begin{equation}
f=\exp\left(\,\,\int\limits_{K_b}^{K_a}\frac{K^2dK}{Z} - \tilde{v}_2\right), 
\quad \int\limits_{K_b}^{K_a}\frac{dK}{Z}=\tilde{v}_0, \quad
\int\limits_{K_b}^{K_a}\frac{KdK}{Z}=\tilde{v}_1
\label{jac2}
\end{equation}
with
\begin{equation}
\tilde{v}_j=v_j - \int\limits_{K_1}^{K_2}\frac{K^jdK}{Z}, \quad j=0,1,2.
\end{equation}
Note that $K_b$ is now on the other sheet of the Riemann surface. Introducing
\begin{equation}
\hat{v}_j=v_j - \Re \int\limits_{K_1}^{K_2}\frac{K^jdK}{Z}, \quad j=0,1,2
\end{equation}
and using the obvious relation
\begin{equation}
\Im \int\limits_{K_1}^{K_2}\frac{K^jdK}{Z} = \frac{1}{4{\rm i}} \left(\,
\oint\limits_{a_1} \frac{K^jdK}{Z} + \oint\limits_{a_2} \frac{K^jdK}{Z}\right)
\end{equation}
we obtain 
\begin{equation}
\tilde{v}_j = \hat{v}_j - \frac{1}{4} \left(\,
\oint\limits_{a_1} \frac{K^jdK}{Z} + \oint\limits_{a_2} \frac{K^jdK}{Z}\right).
\label{vtilde}
\end{equation}
It can easily be verified that the real functions $\hat{v}_j$ are solutions of
the Laplace equation and satisfy the same 
recursion relations (\ref{rec}) as $v_j$.
Note that an asymptotically flat solution ($f\to 1$ at infinity) is obtained
for $v_j\to 0$ (or $\hat{v}_j \to 0$) at infinity. This condition fixes the
integration constants in (\ref{rec}).

In the next section I will discuss physically interesting examples. They differ
in the choice of the potential function $v_0$ (or $\hat{v}_0$) and the 
parameter $K_1$. In all cases I assume 
\begin{equation}
K_2=-\overline{K}_1, \quad \Re K_1 \le 0, \quad \Im K_1 \le 0. 
\label{K12}
\end{equation}   
\section{Examples}
\subsection{The rotating black hole}
The Kerr solution is obtained for real $K_1$, i.e.
\begin{equation}
Z=(K-K_1)(K-K_2)\sqrt{(K+{\rm i}z)(K-{\rm i}\bar{z})},
\end{equation}
and
\begin{equation}
\hat{v}_0=C\left(\frac{1}{r_1}+\frac{1}{r_2}\right)
\label{vv}
\end{equation} 
with a positive parameter $C$ and
\begin{equation}
r_k=\sqrt{(K_k+{\rm i}z)(K_k-{\rm i}\bar{z})}, \quad r_k>0;\,\, k=1,2.
\label{rr}
\end{equation}
{}From (\ref{rec}) we obtain
\begin{equation}
\hat{v}_j=C\left(\frac{K_1^j}{r_1}+\frac{K_2^j}{r_2}\right), \quad  j=1,2.
\end{equation}
The $a$-periods in Eq.\ (\ref{vtilde}) can easily be calculated by means of 
the residues at the poles $K_1$ and $K_2$ [note that $K_2=-K_1$ according
to (\ref{K12})]:
\begin{equation}
\oint\limits_{a_1} \frac{K^jdK}{Z}=\frac{\pi {\rm i}K_1^{j-1}}{r_1}, \quad
\oint\limits_{a_2} \frac{K^jdK}{Z}=\frac{\pi {\rm i}K_2^{j-1}}{r_2}.
\end{equation}
By a suitable combination of Eqs.\ (\ref{jac2}) this leads to
\begin{equation}
f=\exp\left(\,\,\int\limits_{K_b}^{K_a}\frac{dK}
{\sqrt{(K+{\rm i}z)(K-{\rm i}\bar{z})}}\right)
\end{equation}
with
\begin{equation}
\int\limits_{K_b}^{K_a}\frac{dK}
{(K-K_1)\sqrt{(K+{\rm i}z)(K-{\rm i}\bar{z})}} = \frac{1}{r_1}\left(
2CK_1-\frac{\pi {\rm i}}{2}\right),
\end{equation}
\begin{equation}
\int\limits_{K_b}^{K_a}\frac{dK}
{(K+K_1)\sqrt{(K+{\rm i}z)(K-{\rm i}\bar{z})}} = \frac{1}{r_2}\left(
-2CK_1-\frac{\pi {\rm i}}{2}\right).
\end{equation}
These intergrals can elementarily be calculated with the final result
\begin{equation}
f=1-\frac{4M}{r_1+r_2+2M+{\rm i}\frac{J}{\sqrt{M^4-J^2}}(r_1-r_2)}\, ,
\end{equation}
where the parameters $M$ (mass) and $J$ (angular momentum)
are related to $K_1$ and $C$ according to
\begin{equation}
M=K_1\coth (2CK_1), \quad \frac{J}{\sqrt{M^4-J^2}}=\frac{1}{\sinh(-2CK_1)}.
\end{equation}
This is exactly the Ernst potential of the Kerr solution. Note that the 
extreme limit ($J=M^2$) is obtained for $K_1\to 0$ (with $M=1/2C$).
\subsection{The rigidly rotating disk of dust}
The solution describing the gravitational field of 
a rigidly rotating disk of dust (placed at $\zeta=0$, $\varrho\le \varrho_0$)
is obtained for \cite{nm95}
\begin{equation}
K_1=\varrho_0\sqrt{{\frac{{\rm i}-\mu}{\mu}}}, \quad
v_0=\frac{1}{\pi{\rm i}\varrho_o^2}\int\limits_
{-{\rm i}\varrho_o}^{{\rm i}\varrho_o}\frac{D(K)dK}
{\sqrt{(K+{\rm i}z)(K-{\rm i}\bar{z})}}
\label{rig}
\end{equation}
(integration along the imaginary $K$-axis, 
$\Re \sqrt{(K+{\rm i}z)(K-{\rm i}\bar{z})} < 0$ for $\varrho$, $\zeta$ outside
the disk) with
\begin{equation}
D(K)=\frac{\mu\ln\left(\sqrt{1+\mu^2(1+K^2/\varrho_0^2)^2}
+\mu(1+K^2/\varrho_0^2)\right)}{\sqrt{1+\mu^2(1+K^2/\varrho_0^2)^2}}.
\label{D}
\end{equation}
{}From (\ref{rec}) we obtain
\begin{equation}
v_j=\frac{1}{\pi{\rm i}\varrho_o^2}\int\limits_
{-{\rm i}\varrho_o}^{{\rm i}\varrho_o}\frac{D(K)K^jdK}
{\sqrt{(K+{\rm i}z)(K-{\rm i}\bar{z})}}, \quad j=1,2.
\end{equation}
Here $\mu$ is a real parameters. (The total mass $M$ and the angular 
momentum $J$ of the disk 
are functions of $\varrho_0$ and $\mu$.) It turns out that the solution is 
regular for $0\le\mu<\mu_0=4.62966\dots$  where the limit $\mu\to\mu_0$, for
finite $M$, leads to $\varrho_0\to 0$. According to (\ref{rig}), this means 
$K_1\to 0$ and it can be shown that
\begin{equation}
\hat{v}_0\to \frac{1}{Mr}, \quad r=\sqrt{z\bar{z}}=\sqrt{\varrho^2+\zeta^2}.
\end{equation} 
Comparing this with Eqs.\ (\ref{vv}), (\ref{rr}) [note that $r_1$ and $r_2$ 
approach $r$ for $K_1\to 0$], and the final remark of the
previous subsection, we are led to the conclusion that the solution approaches 
exactly
the extreme Kerr metric (for $r>0$) in the limit $\mu\to \mu_0$. More details
concerning this ``ultrarelativistic'' limit can be found in \cite{bw} 
and \cite{m}. 
\subsection{Differentially rotating disks of dust}
\begin{figure}
\centerline{\scalebox{0.85}{\includegraphics{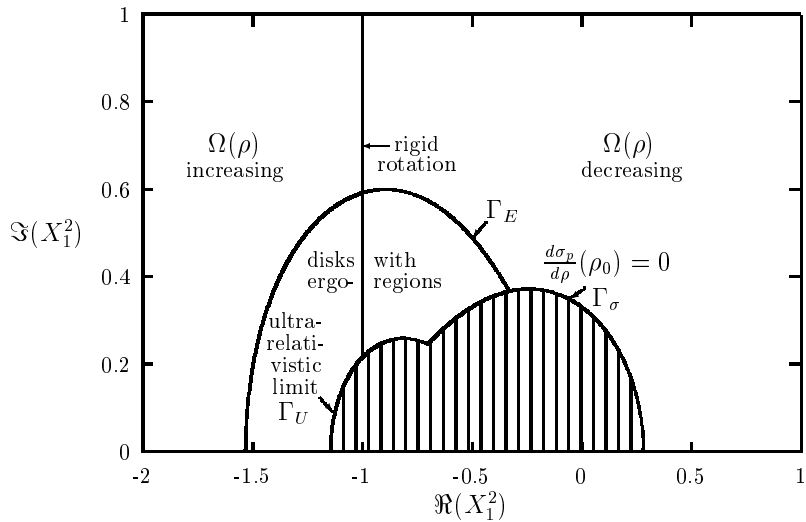}}}
\caption{Differentially rotating disks of dust \cite{ans} 
in dependence on the complex parameter $X_1^2$. Regular solutions have been
found outside the hatched region.}
\label{fig2}
\end{figure}
A three-parameter family of solutions describing differentially rotating disks 
of dust  is obtained for \cite{ans}
\begin{equation}
K_1=\varrho_0 X_1, \quad
v_0=\frac{1}{\pi{\rm i}\varrho_o^2}\int\limits_
{-{\rm i}\varrho_o}^{{\rm i}\varrho_o}\frac{D(K)dK}
{\sqrt{(K+{\rm i}z)(K-{\rm i}\bar{z})}},
\end{equation}
\begin{equation}
\overline{D(K)}=D(-K)=D(K), \, -\varrho_0\le K/{\rm i} \le\varrho_0; \quad
D({\pm\rm i}\varrho_0)=0,
\end{equation}
where $\varrho_0$ is again the (Weyl-coordinate) radius of the disk, 
$X_1$ is an arbitrary 
complex parameter [we only assume $\Re X_1\le 0$, $\Im X_1\le 0$ according
to (\ref{K12})], and $D(K)$ is determined such that the following ``dust 
condition'' \cite{k} is satisfied in the disk, i.e.~for $\zeta=0$, 
$\varrho\le\varrho_0$: 
\begin{equation}
[\Im(A+B-4\varrho AB)]^2 = 4 \,\Im A\, \Im B, \quad
A=\frac{f_{,z}}{f+\bar{f}}\,\, , \, B=\frac{\bar{f}_{,z}}{f+\bar{f}}.
\label{dust}
\end{equation}

Note that for arbitrary $D(K)$ one obtains solutions which might be interpreted
as disks consisting of two counter-rotating streams of particles moving
on geodesics (i.e.~two dust components), see \cite{led}. 
The condition (\ref{dust}) guarantees that there is one stream of particles
only\footnote{In this case the geodesic motion is a consequence of the 
Einstein equations. On the other hand, a formal superposition of two dust 
energy-momentum tensors does not lead automatically 
to a geodesic motion of the particles. 
Therefore, the physical interpretation \cite{klein}
of a particular solution of this class is unsatisfactory.}. The angular
velocity $\Omega(\varrho)$ can be calculated afterwards. 

The solution depends
on the three parameters $\varrho_0$, $\Re X_1$, and $\Im X_1$. According to
(\ref{rig}), the rigidly rotating disk of dust ($\Omega ={\rm const.}$)
is included for $\Re X_1^2 = -1$.
The condition (\ref{dust}) leads to a complicated nonlinear integral
equation for $D(K)$ which has been solved numerically to an extremely high
accuracy, see \cite{ans}. [For $\Re X_1^2 = -1$ one obtains (\ref{D}), of
course.]  A regular solution with positive surface mass-density has been found
in the parameter region as indicated in Fig.\ \ref{fig2}. For $X_1^2$ 
approaching the curve $\Gamma_U$, $\varrho_0\to 0$ follows for finite $M$, and
the extreme Kerr solution is reached again.
This confirms the conjecture formulated by Bardeen and Wagoner \cite{bw} that 
differential rotation will not change the ultrarelativistic 
limit. The curve $\Gamma_{\sigma}$
is characterized by a vanishing derivative of the surface mass--density
$\sigma_p$ at the rim of the disk. (Normally only  $\sigma_p$ itself vanishes 
at the rim.) $\Gamma_E$ divides the parameter-space into parts with and
without ergoregions of the solutions. The angular velocity $\Omega$ is
always a monotonic function of $\varrho$, increasing for $\Re X_1^2<-1$ and
decreasing for $\Re X_1^2>-1$. 

\vspace*{0.25cm} \baselineskip=10pt{\small \noindent I would like to thank
M.~Ansorg, A.~Kleinw\"achter, and G.~Neugebauer for many valuable 
discussions.}

\end{document}